\documentstyle[aps,prd,preprint]{revtex}

\def\beq{\begin{equation}}
\def\eeq{\end{equation}}
\def\bea{\begin{eqnarray}}
\def\eea{\end{eqnarray}}
\def\eqlab#1{\label{eq:#1}}

\def\eref#1{(\ref{eq:#1})}
\def\eqref#1{eq.~(\ref{eq:#1})}
\def\Eqref#1{Eq.~(\ref{eq:#1})}

\def\sla#1{#1 \hspace{-2.4mm} \slash}

\def\boxfrac#1#2{\mbox{\small{$\frac{#1}{#2}$}}}
\def\half{\mbox{\small{$\frac{1}{2}$}}}
\def\thalf{\mbox{\small{$\frac{3}{2}$}}}

\def\third{\mbox{\small{$\frac{1}{3}$}}}

\def\al{\alpha}
\def\be{\beta}
\def\ga{\gamma} \def\Ga{{\it\Gamma}}
\def\de{\delta} \def\De{{\Delta}}
\def\veps{\varepsilon}  \def\eps{\epsilon}

\def\la{\lambda} 
\def\Pit{{\it\Pi}}
\def\Psit{{\it\Psi}}
\def\si{\sigma} 
\def\th{\theta}  \def\Th{\Theta}
  
\def\vphi{\varphi}

\def\dd{{\rm d}}
\def\pa{\partial}

\def\ie{{i.e.\ }}
\def\eg{{e.g.\ }}

\def\pa{\partial}

\def\no{\nonumber}
\def\BK#1#2{{\it #1}, #2}         % Title Publisher
\def\CF#1#2#3#4{#1 {#2} (#3) #4}  % Journal Volume Year Page
\def\ibid {{\it ibid.\ }}

\def\ann {Ann.~Phys.~(NY)}
\def\ijmp {Int.~J.~Mod.~Phys.}
\def\hpa {Helv.~Phys.~Act.}

\def\jpa {J.~Phys.~A}
\def\ncim {Nuovo~Cim.}
\def\np {Nucl.~Phys.}
\def\prev {Phys.~Rev.}
\def\prc {Phys.~Rev.~C}
\def\prd {Phys.~Rev.~D}
\def\plb {Phys.~Lett.}
\def\prl {Phys.~Rev.~Lett.}
\def\prep {Phys.~Rep.}

\def\ptp {Prog.~Theor.~Phys.}

\def\thmp {Theor.~Math.~Phys.}
\def\zp {Z.~Phys.}
\def\rg{{\rm g}}

\def\DD{{\cal D}}
\def\lag{{\cal L}}
\def\ham{{\cal H}}
\def\psib{\bar{\psi}}
\def\Psib{\bar{\Psit}}

\def\DOF{d.o.f.}

\begin{document}

\draft
\preprint{THU-98/07}

\title{Quantization of an interacting spin-3/2 field\\
 and the $\De$--isobar}
\author{V.~Pascalutsa}
\address{Institute for Theoretical Physics, University of Utrecht,
Princetonplein 5,\\
3584 CC Utrecht, The Netherlands}
\date{September 14, 1998}
\maketitle
\thispagestyle{empty}
\begin{abstract}
Quantization of the free and interacting Rarita-Schwinger field is considered
using the Hamiltonian path-integral formulation. The particular interaction
we study in detail is the $\pi N \De$ coupling used in the phenomenology
of the pion-nucleon and nucleon-nucleon systems.
Within the Dirac constraint analysis, we show that there is an excess of
degrees of freedom in the model, as well as the inconsistency related to 
the Johnson-Sudarshan-Velo-Zwanzinger problem. It is further suggested
that couplings invariant under the gauge transformation of the 
Rarita-Schwinger field are generally free from these inconsistencies.
We then construct and briefly analyse some lowest in derivatives
gauge-invariant $\pi N \De$ couplings.
\end{abstract}

\section{Introduction}
A covariant description of the interacting
spin-3/2 field is famous for its various problems and paradoxes.
Presently, supergravity
is the only example of a local field theory which includes a massless 
spin-3/2 field (gravitino) in a consistent way, for a review see
Ref.~\cite{sugra}.
For the particle phenomenology, however,
it would be desirable to construct a consistent description
in a flat space. Such description is needed,
for example, for the treatment of the spin-3/2 baryon resonances, 
like the $\De$(1232)-isobar, in
the low-energy hadron scattering \cite{Nat71,ela,Dav89,nn,memo,chipt}.
Another interesting application is the search for the spin-3/2 leptons
 \cite{ALS96}. 

The major problems in the local\footnote{For non-local formulations see \cite{nlo}.}
 higher-spin field theory
are closely related to the presence of unphysical lower-spin components
in the covariant representation of the field.
More specifically, a field with a given spin $s\ge 1$ , 
besides the physical components, necessarily contains components
of spin $(s-1)$, $(s-2)$, etc. 
For instance, in the Rarita-Schwinger (RS) formalism \cite{RS41} adopted
in this work, the spin-3/2 field is represented by a 16 component 
vector-spinor $\psi_\mu$, while only 4 components are needed for 
the description of a massive spin-3/2 particle and thus the rest
of the components should be attributed to the lower-spin sector.
The free action of such theories is then
constructed in such a way that 
at the level of the equations of motions
the constraints are
produced reducing the number of independent components
to the necessary value (equal to $2s+1$ for a massive and 2 
for a massless particle with spin).

In the interacting case the situation is
generally more complex, since all the components may couple in a non-trivial
way. The constraints are then altered, moreover their amount may
change. In the latter case, \ie if the number of constraints in the
free and interacting theory is different, one can conclude that a wrong number
of degrees of freedom
(\DOF) is interacting, and therefore, this form of
interaction is physically unacceptable. 
Another type of inconsistency which may often arise is
the presence of the famous Johnson-Sudarshan (JS) \cite{JS61} 
and Velo-Zwanzinger (VZ) \cite{VZ69} problems. More recently,
it was shown that JS and VZ problems have a common origin \cite{jsvz}, 
and furthermore they are related to the 
mentioned problem of the constraint violation\cite{hek79,Cox89}.

All these problems are known \cite{Nat71,Hag71,Sin73} to be present for
the coupling of a massive RS field $\psi_\mu(x)$
to a spinor $\Psit(x)$ and a (pseudo-) scalar $\phi(x)$ described by the
the following Lagrangian\footnote{
The conventions used throughout this paper are:
$\hbar=c=1,\, \rg_{\mu\nu}={\rm diag(1,-1,-1,-1)},
\, \veps^{0123}=1,\,\gamma_5=i\ga_0\ga_1\ga_2\ga_3,\, \si_{\mu\nu}=
\half\left[\ga_\mu,\ga_\nu\right],$ spinor indices are usually omitted.},
\beq
\eqlab{int}
\lag_{\rm int}= g\psib_\mu (\rg^{\mu\nu} + a \ga^\mu\ga^\nu)\Psit\,\pa_\nu\phi
+ {\rm H.c.}
\eeq
where $g$ is the coupling constant, and $a$ is related to 
the {\it off-shell parameter}
 $z$ as follows, $a=-z-\half$, cf.\ Ref.\ \cite{Nat71}.
Up to the isospin complications, this interaction represents the 
$\pi N\De$-coupling, frequently used in 
various field-theoretical models of the low-energy 
$\pi N$ and $NN$ interaction\footnote{
For some applications to the $\pi N$ system see \eg Refs.\ \cite{Nat71,ela}
(effective chiral Lagrangians),  \cite{memo} (relativistic 
meson-exchange models), \cite{chipt} 
(chiral perturbation theory), see also Ref.\ \cite{Dav89} for a 
list of common problems in the treatment of the $\De$.}. 

This coupling is also known to have the above mentioned bad property
of involving the unphysical spin-1/2 components. The contribution
of the spin-1/2 sector exhibits itself in the $\De$-exchange
scattering amplitudes
as a substantial spin-1/2 background in addition to the spin-3/2
resonance behaviour around the $\De$ mass position.

In present work, the pathologies of this coupling 
are analysed within the Dirac-Faddeev (DF) quantization framework
\cite{Dir64,Fad70,Fra73,Sen76,GiT86,HeT92,FaS90}. Thus,
first we shall transit to the Hamiltonian formulation, 
find the constraints in the phase-space of the theory
using the Dirac's method\cite{Dir64}
and check whether the above mentioned \DOF\ counting  is 
consistent. Secondly, we shall write down the phase-space path integral
taking the constraints into account, following a generalization \cite{Fra73}
of Faddeev's approach \cite{Fad70}. 
It is usually possible to integrate out the conjugate momenta 
and thus obtain the configuration-space path integral.
The obtained path integral can in principle be different from the 
one we would naively write down without taking the constraints into
account. In this case the naive Feynman
rules (which one would just `read off' the original Lagrangian)
are generally not applicable. 
Applying this procedure to interaction \eref{int}, 
indeed leads to a result different
from the naive one, see \Eqref{pi2}.
On the way to this result, we shall meet the inconsistencies
at the classical level found before using different methods
\cite{Nat71,Hag71,Sin73}.

The question arises whether it is possible in principle to formulate
a consistent interaction of the RS field without supersymmetry, 
or coupling to gravity, or both. As will be argued in section IV,
it is generally possible, if the interaction in question is symmetric
under the gauge transformation of the RS field. In particular, we
construct the following  gauge-invariant $\pi N\De$ coupling,
$$
\lag^{\rm (GI)}_{\rm int}= g\, \veps^{\mu\nu\al\be}\,
(\pa_\mu\psib_\nu) \ga_5\ga_\al \Psit \pa_\be\phi 
+ {\rm H.c.}
$$
which is shown to admit consistent path-integral quantization.
The good properties of this interaction are especially clearly seen
from the tree-level $\De$-particle exchange amplitude, \Eqref{tree2}.
It is proportional to the spin-3/2 projection operator, and,
at the same time, is not singular at $p^2=0$. Consequently, the decoupling 
of the spin-1/2 component of the RS field is achieved in the
manifestly covariant and local formulation. The spin-1/2 
background is absent.

The paper is organized as follows.
In the next section we work out the DF
procedure for the free 
massive spin-3/2 field. This discussion  serves
mainly as an introduction to the formalism. In section III
we perform the Dirac constraint analysis
of the conventional $\pi N\De$ interaction 
\eref{int}, notify the presence of the JS-VZ problem, and
obtain the configuration-space path integral of the model.
In section IV we argue that gauge-invariant interactions
do not, in general, alter the number of constraints, and consider some
lowest in derivatives gauge-invariant $\pi N \De$ couplings. 
The conclusions
are formulated in section V. Finally, 
an extension of the St\"uckelberg formalism to the case of
the spin-3/2 field is given in the Appendix. 

\section{Free Rarita-Schwinger field}
The quantization of the free RS field in Hamiltonian formulation
was considered previously in Refs.
\cite{FrV77,Sen77,BaT78,EnK79}. In this section we shall briefly recapitulate
these considerations in order to summarize the results
and set up the framework. 
Also, the free-field quantization is usually done on Majorana (Hermitian)
field, while here we work with the complex field, hence allowing for the
charge. This leads only to minor modifications related to
the doubling of the field components and corresponding
\DOF\ and constraints.

The free Lagrangian of a complex RS field $\psi_\mu(x)$ with mass $m$ is
written as follows, 
\bea
\eqlab{l32}
\lag_{\small 3/2}&=& \half \bar{\psi}_\mu\, \{ \si^{\mu\nu},\,(i\sla{\pa}-m)\}\,\psi_\nu
= -\half \veps^{\mu\nu\al\be}\,
\psib_\mu\ga_5\ga_\al\pa_\be\psi_\nu \no\\ 
&&+\half \veps^{\mu\nu\al\be}\,
(\pa_\be\psib_\mu)\ga_5\ga_\al\psi_\nu - m\psib_\mu \si^{\mu\nu}\psi_\nu.
\eea
To determine the constraints we follow the path of Dirac\cite{Dir64}.
From the definition of conjugate momenta,
\bea
\pi^{\mu\dagger}(x)&=& {\pa\lag / \pa \dot{\psi}_{\mu}(x)},\,\,\,
\pi^{\mu}(x) = {\pa\lag / \pa\dot{\psi}^{\dagger}_{\mu}(x)}, \no
\eea
we find the following primary constraints:
\bea
\eqlab{freepconstr}
\th_0(x)&=& \pi_0(x),\,\,\,
\th_i(x) =\pi_i(x) + \half \veps_{ijk}\ga_0\ga_5\ga_k\psi_j(x)\\
\th_0^{\dagger}(x)&=& \pi_0^{\dagger}(x),\,\,\,
\th_i^{\dagger}(x) =\pi_i^{\dagger}(x)
 + \half \veps_{ijk}\psi_j^{\dagger}(x)\,\ga_0\ga_5\ga_k. \no
\eea
The Hamiltonian, $H=\int\!\dd^3x\,\ham(x)$, is then given by
\beq
\eqlab{ham32}
\ham_{\small 3/2} = \left[\bar{\psi}_i (\veps_{ijk} \ga_5\ga_j\,\pa_k
 - m \bar{\psi}_i \ga_i\ga_0 ) \psi_0 + {\rm H.c.}\right] 
+ \psib_i (\veps_{ijk}\, \ga_0\ga_5\,\pa_k + m \si_{ij}) \psi_j
\eeq
We also introduce the fundamental Poisson brackets (defined at $x_0=y_0$):
\beq
\{\psi_{\mu\si} (x), \,\pi_\tau^{\nu \dagger}(y)\}_P
  =  \de^\nu_\mu\,\de_{\si\tau}\,\de^3(x-y), 
\eeq
here we have written out the spinor indices $\si,\tau=0,\ldots,3$.
In the following we will omit them again. Brackets 
involving only fields or only momenta vanish.\footnote{
From the property of the Poisson bracket,
$\{ A, B\}_P^{\dagger} = -\{B^\dagger,A^\dagger\}_P,$
we have
$\{\psi_\mu^\dagger (x), \,\pi^{\nu}(y)\}_P
  =  - \de^\nu_\mu\,\de^3(x-y).$}

The primary constraints should now be added to the Hamiltonian through
the Lagrange multipliers to form the {\it total Hamiltonian} density:
\beq
\ham_T= \ham_{\small 3/2} + \la_0 \th_0 + \la_i \th_i + {\rm H.c.}
\eeq
To guarantee the conservation of constraints in time one requires
that they commute with the total Hamiltonian, \ie the corresponding
Poisson bracket must vanish. 

From condition
$\{\th_i(x),\,H_T\}_P=0,$ the Lagrange multipliers $\la_i$
can be determined. Constraints $\th_i$ are thus second class and we
may resolve them right away by introducing the Dirac bracket:
\bea 
\eqlab{diracbra}
\{ A(x), B(y)\}_D &=& \{A(x),\,B(y)\}_P 
-\int\!\dd^3 z_1\, \dd^3 z_2\, \{A(x),\th_i^\dagger(z_1)\}_P \no\\
&&\times\left(\{\th_i^\dagger(z_1),\th_j(z_2)\}_P
\right)^{-1} \{\th_j(z_2),B(y)\}_P
\eea
To this end we can find,
\bea
\{\th_i(x),\th_j^\dagger (y)\}_P&=&
 -i \si_{ij} \de^3(x-y), \\
\left(\{\th_i(x),\th_j^\dagger (y)\}_P\right)^{-1}
&=& -\half i\ga_j\ga_i\,\de^3(x-y),
\eea 
hence\footnote{One can get to this and some other results in a
more efficient way by using the Hamiltonian reduction \cite{FaJ88} instead of
Dirac's analysis. (We thank L.D. Faddeev for this remark).}
\beq
\eqlab{dbramass}
\{\psi_i(x),\psi^\dagger_j(y)\}_D = \half i \ga_j\ga_i\,\de^3(x-y). 
\eeq

From the condition that $\th_0$ and $\th_0^\dagger$ commute with
the (total) Hamiltonian we find the secondary constraints\footnote{Note the identities: 
$\veps_{ijk}\ga_5\ga_k = -i\si_{ij}\ga_0,\,$
$\half i\veps_{ijk}\ga_j\ga_k=\ga_5\ga_0\ga_i,\,$
$\half \veps_{ijk}\veps_{lmn} \ga_j\ga_m\ga_k\ga_n = -\si_{il}.$},
\bea
\th_4(x)&=& -i \si_{ij}\pa_i\psi_j + m \ga_i\psi_i \no\\
\th_4^\dagger(x) &=& - i\pa_i\psi_j^\dagger \si_{ij}
- m \psi_i^\dagger\ga_i.
\eea
We may rewrite the Hamiltonian in the following fashion,
\beq
\ham_{\small 3/2} =  \th_4^\dagger\psi_0 + \psi_0^\dagger \th_4  
+ \psib_i (\veps_{ijk}\, \ga_0\ga_5\,\pa_k + m \si_{ij}) \psi_j
\eeq
Now one can immediately see that the tertial constraints $\th_5$
(and $\th_5^\dagger$) arising from $\{\th_4(x), H_T\}_D=0$
(and $\{\th_4^\dagger(x), H_T\}_D=0$) are linear in
 $\psi_0$ ($\psi_0^\dagger$) with the following
proportionality coefficient,
\beq
\eqlab{rfree}
\int\!\dd^3x\,\{\th_4(x), \th_4^\dagger(y) \}_D = \thalf i m^2.
\eeq
Clearly, the conditions that $\th_5$ and $\th_5^\dagger$
commute with the total Hamiltonian determine
the remaining Lagrange multipliers $\la_0$ and $\la_0^\dagger$, thus
no more constraints arise. It is also clear that all the 
constraints are second class.

We can perform now an exercise in the \DOF\ counting.
The field $\psi_\mu$ and its conjugate momentum $\pi^\mu$ have 
$4\times 4=16$ (complex) components each, so 32 in total. We have
$6\times 4=24 $ (complex) constraints on them. Hence the
number of independent components is 8: precisely what is needed
for the description of the spin \DOF\ in the phase-space
of a massive spin-3/2 particle.

In the massless case the situation is somewhat different.
The requirement 
$$\{\th_4(x),H_T\}_D=0$$ 
becomes an identity,
and no $\th_5$ constraints arise. We then have only 5 fermionic
constraints, where $\th_i$ are second class while $\th_0$ and $\th_4$
are first class. The appearance of the first-class constraints is, of course,
related to the fact that the massless Lagrangian is (upto a total
derivative) invariant under
the gauge transformation, 
\beq
\eqlab{gauget}
\psi_\mu \rightarrow \psi_\mu + \pa_\mu \eps,
\eeq
where $\eps(x)$ is a complex fermionic field.
To each first-class constraint we have to introduce a gauge-fixing condition.
The \DOF\ counting is then also consistent: we are left with 4 independent
field components in the phase-space which is appropriate for a massless
particle with spin.

Let us now proceed to the path-integral quantization of the system.
We concentrate on the massive case. Following the generalization of 
Faddeev's procedure \cite{Fad70} to the case of (fermionic) second-class
constraints \cite{Fra73,Sen76,GiT86,HeT92} we write
down the phase-space path integral in the following form,
\bea
\eqlab{path}
Z&=&\int\!\DD\psi_\mu\,\DD\psi_\mu^\dagger\,\DD\pi^\mu
\,\DD\pi^{\mu\dagger}
\,\left(\det{\|\{\th,\th\}_P\| }\right)^{1/2} \,\prod_{n=0}^{5} \de(\th_n)
\, \de(\th_n^\dagger)\no \\
&&\times \exp{\left\{i\int\!\dd^4 x\left[ \pi^{\mu\dagger}\dot{\psi}_\mu
+ \dot{\psi}_\mu^\dagger\pi^\mu - \ham_{3/2}
\right]\right\} },
\eea
where $\|\{\th,\th\}_P\|$ represents the matrix of Poisson brackets of 
constraints. In our case it is,\footnote{Note that nowhere our calculation 
do we need to know fully  
$\th_5$ constraint. It suffices to know that $\th_5$ is linear in $\psi_0$
with the already determined coefficient $\thalf im^2.$ This observation
has been made also in Ref.\ \cite{Yam86}.}
\beq
\left( \begin{array}{cc}
0 & \|\{\th(x),\th^\dagger(y)\}_P\| \\
\|\{\th^\dagger(x),\th(y)\}_P\| & 0 
\end{array} \right)
\eeq
where
\beq
\|\{\th(x),\th^\dagger(y)\}_P\| = \left( \begin{array}{cccc}
0 & 0 & 0 & \thalf i m^2 \\
0 & -i\si_{ij}& \Ga_i & \{\th_i, \th_5^\dagger\} \\
0 & \Ga_j & 0 & \{\th_4, \th_5^\dagger\} \\
\thalf i m^2 & \{\th_5, \th_j^\dagger\} &
\{\th_5, \th_4^\dagger\} & \{\th_5,\th_5^\dagger
\}
\end{array} \right) \de^3(x-y),
\eeq
$$\Ga_i \equiv i \si_{ij}\pa_j + m \ga_i
$$
The calculation of the determinant and integration over $\pi$'s produce
the following result,
\beq
\eqlab{pi1}
Z=\int\!\DD\psi_\mu\,\DD\psi_\mu^\dagger\,
\det{[ (i \ga_i\pa_i + \boxfrac{3}{4} m)\,\de^3(x-y)]}\,e^{i\int \lag_{3/2}}. 
\eeq
The determinant is field-independent and can be dropped, we
have kept it just for further comparison to the interacting case. Having 
obtained path integral \eref{pi1} we complete
the DF quantization
of the free massive spin-3/2 field and conclude that constraints
do not modify the original Lagrangian, hence the `naive' Feynman rules apply.

We will not treat separately the massless case (this is done in details in 
Refs.\ \cite{FrV77,Sen77}). Instead, we may apply an analog of the
St\"uckelberg mechanism\cite{Stu57}, which 
allows us to treat the massless and
massive case on the same footing. This analysis is done in the Appendix.

\section{The $\pi N \De$--coupling model}
In this section we apply the Dirac--Faddeev procedure to quantize 
the $\pi N \De$ phenomenological interaction discussed in Introduction.
The model is given by the following Lagrangian,
\bea
\eqlab{pin}
\lag &=& \lag_0 + \lag_{\small 1/2} + \lag_{\small 3/2} + \lag_{\small\rm int},
\\
\lag_0 & = & \half \pa_\mu \phi\,\pa^\mu\phi - \half \mu^2\phi^2,\no\\
\lag_{1/2}&=& \Psib (i\sla{\pa} -M )\Psit=
\half i \Psib \ga_\mu\pa_\mu \Psit - 
\half i(\pa_\mu\Psib) \ga_\mu \Psit - M \Psib\Psit, \no
\eea
where $\lag_{\small\rm int}$ and $\lag_{\small 3/2}$ are defined in \Eqref{int}
and \eref{l32} respectively. 

We follow precisely the same steps as in the preceding section.
In addition to $\pi^\mu$ we define
\bea
P(x) &=& {\pa\lag / \pa \dot{\phi}(x) }, \no\\
\Pit^{\dagger}(x) &=& {\pa\lag / \pa\dot{\Psit}(x)},\,\,\,
\Pit(x) = {\pa \lag / \pa\dot{\Psit}^{\dagger}(x)} \eea
and find the `velocity' $\dot{\phi}$:
\bea
\eqlab{velocity}
\dot{\phi}(x)&=&P(x)-F[\Psit(x),\psi_\mu(x)], \\
F[\Psit,\psi_\mu]&\equiv & g (1+a) \psib_0\Psit -ga \psib_i\ga_i\ga_0\Psit
+{\rm H.c.}, \no
\eea
and the following primary constraints (in addition to \Eqref{freepconstr}):
\beq
\eqlab{pconstr}
\chi(x)= \Pit(x) - \half i\Psit(x),\,\,\,
\chi^{\dagger}(x) = \Pit^{\dagger}(x) + \half i\Psit^{\dagger}(x). 
\eeq
The model Hamiltonian is given by,
\bea
\eqlab{ham}
 \ham &=& \ham_0 + \ham_{1/2} + \ham_{3/2} + \ham_{\rm int},\\
\ham_0&=& \half (P^2 - F^2) + \half (\pa_i \phi)^2 + \half\mu^2\phi^2, \no\\
\ham_{1/2}&=&\bar{\Psit}(i\ga_i\pa_i + M)\Psit, \no \\
\ham_{\rm int}&=& -\lag_{\rm int} = 
-(P-F)\, F + g \left[ a\psib_0\ga_0\ga_i\pa_i\phi + \psib_i
(\de_{ij} - a \ga_i\ga_j)\pa_j\phi\right] \Psit + {\rm H.c.} \no
\eea
with $\ham_{3/2}$ given in \Eqref{ham32}.

We postulate the fundamental Poisson brackets,
\bea
\{\phi(x), P(y)\}_P &=& \de^3(x-y),\no\\
\{\Psit_\si (x), \Pit_\tau^{\dagger} (y)\}_P &=& \de_{\si\tau}\,\de^3(x-y), \\
\{\psi_{\mu\si} (x), \,\pi_\tau^{\nu \dagger}(y)\}_P
  &=&  \de^\nu_\mu\,\de_{\si\tau}\,\de^3(x-y), \no
\eea
all the other brackets vanish. Note that the brackets
are symmetric in the case of fermionic variables (such as
$\Psit,\Pit,\psi_\mu,\pi^\mu$) and anti-symmetric in the case of 
bosonic variables (such as $\phi,P,\ham$), they are also
anti-symmetric in the mixed case.

Next we resolve the second-class constraints $\th_i$ and $\chi$
by introducing corresponding Dirac brackets, and note
\beq
\eqlab{dbranucl}
\{\Psit(x),\Psit^\dagger(y)\}_D =  -i \de^3(x-y)
\eeq
while $\{\psi_i(x),\psi^\dagger_j(y)\}_D$ is given by \Eqref{dbramass}. 

A crucial point here is that the 
condition of conservation of $\th_0$
constraint leads to a constraint which in general contains $\psi_0$.
Namely,
\beq
\th_4 =   -i \si_{ij}\pa_i\psi_j
+ m \ga_i\psi_i - ag\ga_i\Psit\pa_i\phi +g(1+a)\,(P-F)\,\ga_0\Psit
\eeq
and similar for $\th_4^\dagger$. It is $F$ that has an explicit dependence
on $\psi_0$ as given by \Eqref{velocity}.

As we saw in the previous section, the constraint containing  $\psi_0$
is always the last one in the chain of constraints. Hence for $a\neq -1$,
$\th_4$ is the last constraint, and we have then 5 ($\times 4$) constraints,
all of them being second-class.
Counting the number of \DOF\ for this case we certainly 
find an excess of them, because we 
are one constraint too short as compare to the free case where the \DOF\ 
counting is built in correctly. Thus, we conclude that for $a\neq -1$ the $\pi N\De$ interaction considered here is {\it inconsistent} with the free theory construction. The same conclusion has been drawn by Nath, Etemadi and Kimel \cite{Nat71} based on a constraint analysis in Lagrangian formulation.
The choice $a=-1$ is thus preferable
and we continue the analysis for this case only.

For $a=-1$, the $\th_4$ constraints read
\bea
\th_4(x)&=& -i \si_{ij}\pa_i\psi_j + m \ga_i\psi_i + g\ga_i\Psit\pa_i\phi\no\\
\th_4^\dagger(x) &=& - i\pa_i\psi_j^\dagger \si_{ij}
- m \psi_i^\dagger\ga_i - g\Psit^\dagger\ga_i\pa_i\phi.
\eea
As in the free case, constraints $\th_5$ and $\th_5^\dagger$
are linearly proportional to $\psi_0$. Now only with a different
coefficient:
\beq
\eqlab{rfactor}
R(x)\equiv\int\!\dd^3y\,\{\th_4(x), \th_4^\dagger(y) \}_D 
= i\left[\thalf m^2 - g^2 (\pa_i\phi)^2
\right].
\eeq
At this point we hit another problem. The coefficient
may vanish when $\thalf m^2 = g^2 (\pa_i\phi)^2$.
Then, either the $\th_4$ constraints are first-class, or
we will find some further second-class constraints. In any case the
\DOF\ counting will again be different from that of the free theory. In the
massless case the situation is even worse since the problem occurs for
any value of $g^2 (\pa_i\phi)^2$.

It is interesting to note that the same problem arises
in the constraint analysis of the minimal
coupling of the RS field to the external electromagnetic field 
\cite{hek79,Cox89}. There it was identified with the JS-VZ problem.
On the other hand, Hagen \cite{Hag71} and Singh \cite{Sin73} 
revealed the JS and VZ problems 
in the $\pi N\De$ coupling being considered. Their analysis
is done in lines with the original treatment \cite{JS61,VZ69}
and thus is rather different from ours; nevertheless, the factor
giving rise to the JS and VZ problem in their works is
precisely $R(x)$ of \Eqref{rfactor}. Moreover, we can easily compute the
field commutators taking into account $\th_4$ constraints
(\ie the second stage Dirac bracket),
and find that the corresponding quantum commutators are not positive-definite, because $R$ is not, in line with Hagen's conclusion.
We can therefore confirm the observation of \cite{jsvz,hek79} that the JS-VZ
problem appears itself in the violation of constraints. 

To proceed with the quantization 
let us assume $R(x)\neq 0$ (although note that this is not
a Lorentz-invariant condition), and write down the
path integral. According to \Eqref{path} we need,
\beq
\left(\det{\|\{\th,\th\}_P\| }\right)^{1/2} = \det\left( \begin{array}{ccccc}
0 & 0 & 0 & R & 0\\
0 & -i\si_{ij}& \Ga_i & \{\th_i, \th_5^\dagger\} & 0\\
0 & \Ga_j & 0 & \{\th_4, \th_5^\dagger\} & g \ga_i\pa_i\phi \\
R & \{\th_5, \th_j^\dagger\} & \{\th_5, \th_4^\dagger\} & \{\th_5,\th_5^\dagger
\}& \{\th_5, \chi^\dagger\} \\
0& 0 & g \ga_i\pa_i\phi & \{\chi, \th_5^\dagger\}& -i 
\end{array} \right) \de^3(x-y)
\eeq
Simplifying this determinant and carrying out the integration over the
conjugate momenta we obtain
\beq
\eqlab{pi2}
Z=\int\!\DD\psi_\mu\,\DD\psi_\mu^\dagger\,
\DD\Psit\,\DD\Psit^\dagger\,\DD\phi\,
\det{\left[ \left(i \ga_i\pa_i + \boxfrac{3}{4} m - \frac{g^2}{2m}\, 
(\pa_i\phi)^2 \right)\,\de^3(x-y)\right]}\,e^{i\!\int\!\lag}. 
\eeq
Thus, our final path integral differs from the naive path integral
by the non-trivial determinant entering the measure.

Non-covariant field-dependent determinants do often occur in the
Hamiltonian path-integral quantization of systems with second-class
constraints, see Refs.~\cite{FVi73,HeS94}.  Usually their contributions
to the Green functions is cancelled by the singular terms coming from the
time-ordering operators, so that resulting Green functions are covariant.
It would be interesting to see whether this mechanism occurs also in
the case of \Eqref{pi2}, or, perhaps, there is indeed some breaking of
Lorentz symmetry suggested by the presence of the JS-VZ problem.

\section{Gauge-invariant couplings}
In the previous section we have seen that the conventional 
$\pi N \De$ interaction suffers from inconsistencies
related to the violation of constraints, in particular 
the JS-VZ problem.
On the other hand, it is intuitively clear that 
(i) gauge-invariant couplings are generally
consistent with the \DOF\ counting. 
Indeed, the number of constraints
is  related to the number of local symmetries of the Lagrangian,
while gauge-invariant couplings do not destroy the symmetry of the
free RS Lagrangian where the \DOF\ counting is correct. We can 
prove statement (i) more rigorously for the linear couplings 
of the RS field, \ie the case when the interaction Lagrangian is given by
\beq
\eqlab{l32int}
\lag_{\rm linear} = \psi_\mu^\dagger J^{\mu} + {\rm H.c.},
\eeq
$J^{\mu}$ is independent of $\psi_\mu$
The proof proceeds as follows 
(we basically follow the proof of equation (8.2.5) in 
Ref.~\cite{Wei95}).

If other fields do not change under the gauge transformation,
we can concentrate just on the $\psi_\mu$ dependent part of
the Lagrangian, which is
$$
\lag = \lag_{3/2}+\lag_{\rm linear}.
$$
The gauge-invariance of the massless Lagrangian then implies
\beq
\eqlab{conserve}
\pa_\mu J^\mu=0.
\eeq
Determining the constraints, we find the usual primary constraints,
\beq
\th_0=\frac{\pa \lag}{\pa (\pa_0\psi_0)}
\eeq
and $\th_i$ of \Eqref{freepconstr}.
The  $\th_i$ constraints do not produce any 
secondary constraints, while requiring time-independence
of $\th_0$ gives us the usual 
\beq
\th_4=\pa_0\th_0.
\eeq
Now, using the Euler-Lagrange field equations and \Eqref{conserve},
we obtain 
\beq
\pa_0\th_4= m \si^{\mu\nu}\pa_\mu\psi_\nu.
\eeq 
If the field is massless, $m=0$, then the time constancy of $\th_4$ is
trivially obeyed and no more constraints arise. If $m\neq 0$, then
we obtain the usual, for the massive theory, second-class constraint,
$\th_5\equiv\pa_0\th_4$.
Thus, only the mass term can 
affect the number of constraints and DOF, 
which proves (i) for the case of linear
coupling.

According to (i) it seems promising to 
search for consistent $\pi N \De$ couplings among the gauge-invariant 
ones. The simplest way to construct those
is to couple the RS field to an explicitly conserved current.
(Actually, the only other way we can see is to allow the pion and the 
nucleon field also transform under the gauge transformation, similarly
to how they transform under the photon gauge transformation. This, however,
would obviously require a supersymmetric realization.
Although an interesting possibility, here we restrict ourselves
to non-supersymmetric realizations.)
 
The lowest in derivatives explicitly gauge-invariant 
$\pi N \De$ interaction is given by 
the following Lagrangian,
\beq
\eqlab{int2}
\lag_{\rm int}= g (\pa_\mu\psib_\nu) \si^{\mu\nu} \Psit \phi
+ {\rm H.c.}
\eeq
However, this interaction is in some sense trivial: 
it describes the coupling of the nucleon and pion to $\pa\cdot\psi$
and $\ga\cdot\psi$, \ie the spin-1/2 sector of the $\De$ field. 
Furthermore, the corresponding tree-level Feynman amplitude for the $\pi N$ scattering  through a virtual $\De$ exchange, 
\beq
\eqlab{ample}
M(p) = \Ga^\al(p) S_{\al\be} (p) \Ga^\be(p)
\eeq
where $p$ is 4-momentum of the $\De$, $\Ga^\al(p)$ and $S_{\al\be} (p)$
are the naive Feynman rules for the vertex and the RS propagator respectively,
\bea
\Ga^\al(p) &=& g\, \si^{\al\mu} p_\mu \\
\eqlab{RS1}
S_{\alpha \beta}(p)&=&\frac{\sla{p} + m}{p^2 - m^2} \left[ 
\rg_{\alpha\beta} - \third\gamma_{\alpha}\gamma_{\beta}
       - \frac{1}{3m^2}(\sla{p}\gamma_\alpha p_\beta + p_\alpha\gamma_\beta
\sla{p})\right],
\eea
vanishes exactly: $M(p)=0$, for all $p$. Having such a classically
`invisible' $\De$ is maybe interesting in some scenarios, but certainly
not in the applications we are interested in here.
We thus should conclude that the $\pi N \De$ 
interaction \eref{int2} involves a correct number of
 $\De$'s field components, however, they have wrong spin representing
parts of the spin-1/2 sector of the RS field, consequently this
interaction can not describe a physical coupling to the spin-3/2 particle.

The next lowest in derivatives gauge-invariant interaction is
written down in the Introduction, and reads as follows 
\beq
\eqlab{int3}
\lag_{\rm int}= g\, \veps^{\mu\nu\al\be}\,
(\pa_\mu\psib_\nu) \ga_5\ga_\al \Psit \pa_\be\phi 
+ {\rm H.c.}
\eeq
For this interaction the tree-level amplitude does 
not vanish. Moreover the result is not sensitive to
$1/m^2$ term of the RS propagator, thus a well-defined massless
limit is guaranteed. We shall
discuss the tree-level calculation in more detail, but first
let us perform the DF quantization of this interaction.

To treat the massive and massless case simultaneously
we introduce the St\"uckelberg spinor $\xi(x)$ described in Appendix.
Our model Lagrangian is thus defined by Eqs.~\eref{pin}, \eref{stuckl32} and
\eref{int3}.

The model has the following primary and secondary
constraints (the hermitian conjugates are omitted),
\bea
\eqlab{giconstr}
\th_i &=& \pi_i - i \si_{ij} (\half\psi_j+ g\Psit\pa_j\phi),\no\\
\th_S &=&\eta - m\ga_i\psi_i, \no\\
\chi &=& \Pit - \half i\Psit, \\
\th_0 &=& \pi_0, \no\\
\th_4 &=& -i \si_{ij}\pa_i\psi_j + m \ga_i\psi_i
 + m \ga_i\pa_i\xi -ig\si_{ij}\pa_i\Psit\pa_j\phi, \no
\eea
and the Hamiltonian density given by, 
\bea
\ham &=&  \half (P-F)^2 + \half (\pa_i \phi)^2 + \half(\mu\phi)^2 
+\bar{\Psit}(i\ga_i\pa_i + M)\Psit\no\\
&&+\left[
 \psi_0^\dagger \th_4  
+ \half \psib_i (\veps_{ijk}\, \ga_0\ga_5\,\pa_k + m \si_{ij}) \psi_j
+ m\psib_i \si_{ij}\pa_j\xi - g\,\veps_{ijk}\,(\pa_i\psi_j^\dagger)
\ga_5\Psit\pa_k\phi + {\rm H.c.}\right],
\eea
where $F=-ig(\pa_i\psi_j^\dagger)\si_{ij}\Psit+{\rm H.c.}$, and 
$\th_4$ is given in \Eqref{giconstr}.

Once again, we introduce the Dirac bracket with respect 
to the second-class constraints
$(\th_i,\,\th_S,\,\chi)$
and find that the field commutators remain to be
given by Eqs.~\eref{dbra}, \eref{dbranucl}.

Now, as can be shown by a direct computation, but also follows from 
the proof given in the beginning of this section, 
the secondary constraint, $\th_4$, commutes
with the total Hamiltonian.  Hence, constraints
\eref{giconstr} are all constraints in the model.

The construction of the path integral goes in exactly the same way as 
discussed in Appendix. Although in this case the matrix of the second-class 
constraint Poisson brackets,
\beq
\|\{\th^{(2)},\th^{\dagger(2)}\}_P\|  = -\left( \begin{array}{ccc}
i\si_{ij}& m\ga_i & ig \si_{ik}\pa_k\phi\\
m\ga_j & 0 & 0 \\
ig \si_{jk}\pa_k\phi & 0 & i 
\end{array} \right) \de^3(x-y),
\eeq
is field-dependent, its determinant is not,
\beq
\det\|\{\th^{(2)},\th^{\dagger(2)}\}_P\|=\det [3im^2\de^3(x-y)]
\eeq
and can be neglected.

Taking the Coulomb gauge, integrating out
the momenta and covarianizing the measure 
we obtain the following configuration-space
path integral of the model,
\beq
Z=\int\!\DD\psi_\mu\,\DD\psi_\mu^\dagger\,
\DD\Psit\,\DD\Psit^\dagger\,\DD\phi\,
\DD\xi\,\DD\xi^\dagger\,
\de(\ga\cdot\psi)\,\de(\psi^\dagger\cdot\ga)\,
e^{i\!\int\!\lag }.
\eeq

Another important simplification which occurs here
due to the gauge symmetry is the decoupling of  
the St\"uckelberg spinor.
We thus may easily integrate it out as well, obtaining
\beq
\eqlab{gipi}
Z=\int\!\DD\psi_\mu\,\DD\psi_\mu^\dagger\,
\DD\Psit\,\DD\Psit^\dagger\,\DD\phi\,
\de(\ga\cdot\psi)\,\de(\psi^\dagger\cdot\ga)\,
\de(\pa\cdot\psi)\,\de(\pa\cdot\psi^\dagger)\,e^{i\!\int\!\lag },
\eeq
where the free spin-3/2 Lagrangian is now given by \Eqref{l32}, while
the rest of the terms in $\lag $ remain unchanged.
Note that starting from the transverse gauge we would obtain 
the same expression.

Let us now reconstruct the Feynman rules for the RS field.
The delta functions in our final path integral clearly indicate
that the Green functions are independent of the spin-1/2 sector
of $\psi_\mu$. We can use for instance the following ``Feynman gauge''
expression for the spin-3/2 propagator
\beq
\eqlab{Feyngauge}
S_{\alpha \beta}(p)=\frac{1}{\sla{p} - m} \left(
\rg_{\alpha\beta}-\third\ga_\al\ga_\be \right).
\eeq
The expression for the vertex reads 
\beq
\Ga^\mu (k,p)= ig\, \veps^{\mu\nu\al\be}
p_\nu \ga_5\ga_\al k_\be,
\eeq
where $k$ is pion momentum, while $p$ can be chosen to represent the 
momentum of either the $\De$ or the nucleon.

Using these rules we can easily compute the
tree-level amplitude for the $\pi N$ scattering through the $s$-
or $u$-channel $\De$
exchange (forgetting about the isospin),
\beq
\eqlab{tree2}
M(k',k;p)=\Ga^\al (k',p)\,S_{\alpha \beta}(p)\,\Ga^\be (k,p) =
\frac{g^2}{\sla{p} - m}\, p^2\, P^{3/2}_{\al\be}(p)\, {k'}^\al k^\be,
\eeq 
where 
\beq
P^{3/2}_{\al\be}(p) = \rg_{\alpha\beta} - \third\gamma_{\alpha}\gamma_{\beta}
       - \frac{1}{3p^2}(\sla{p}\gamma_\alpha p_\beta + 
p_\alpha\gamma_\beta
\sla{p}),
\eeq
is the spin-3/2 projection operator. This operator has the well-known
property of projecting on the spin-3/2 states and is a clear
signature of the spin-3/2 components. Our amplitude
is thus independent of the spin-1/2 sector  of the RS field, which is
certainly the result we desired to obtain. 

The spin-3/2 projection operator was used previously in some phenomenological
models although in a rather {\it ad hoc} way, such as, for example, replacing
the tensor part of the RS propagator by the projection operator, etc.,
 see \eg references cited in \cite{Dav89}. However in these models
problems arise due to the $1/p^2$ non-locality of the projection operator.
In \Eqref{tree2} this problem is obviously not present, which is not surprising
since we depart from a local Lagrangian.

It may look that the JS-VZ problem for coupling \eref{int3} is avoided
just because we made use of the St\"uckelberg mechanism: 
$\th_4$ is then guaranteed to be the first-class constraint and 
the problem discussed below \Eqref{rfactor} can not occur. Suppose, however,
we do not introduce the St\"uckelberg field. In this case, $\th_4$
is given by \Eqref{giconstr} with $\xi=0$, and the commutator of $\th_4$
constraints is given by \Eqref{rfree}, \ie is exactly the same as in the
free theory. Thus, the JS-VZ problem does not occur here, independently
of whether the St\"uckelberg field is used or not.

On the other hand, suppose we would like to avoid the JS-VZ problem
in the conventional coupling by using the St\"uckelberg mechanism.
Then, indeed, the corresponding $\th_4$ constraint becomes first
class, hence its commutator vanishes instead of being field-dependent
as in \Eqref{rfactor}. In that case, however, the St\"uckelberg field
does not ever decouple and the excess of \DOF\ becomes thus explicit,
leading again to the unitarity problem.

\section{Summary and Conclusion}
The Dirac-Faddeev quantization method is very well suited for
analyzing the interacting spin-3/2 field, since it provides a
straight-forward procedure where the control over the degrees of
freedom can be done in a simple transparent way. 
We have applied this procedure to the conventional $\pi N \De$ coupling,
\Eqref{int}, and find this coupling has a number of problems precisely
due to the coupling to extra \DOF. This goes in line with
some previous analyses\cite{Nat71,Hag71,Sin73}, as well as with
the common knowledge that this coupling always produces unphysical spin-1/2
backgrounds in addition to the spin-3/2 contribution. 
For the choice $a=-1$, the problem is not so pronounced, nevertheless
it is present and can be related to the well-known JS-VZ problem.
Furthermore, we argue that for this choice the `naive' Feynman rules 
may be unapplicable since in principle
there are contributions from the determinant in the 
path integral \Eqref{pi2}.

Further, we have suggested to use couplings which are invariant
under the gauge 
transformation of the RS field \Eqref{gauget}. As has been
conjectured and partially proved in section IV, these couplings are
generally consistent with 
the \DOF\ counting (unitarity). We have considered two lowest in derivatives
gauge-invariant $\pi N\De$ couplings. The first one describes the
coupling to purely the spin-1/2 sector of the RS field, and we abandon
its further analysis for this reason. The second coupling,
\Eqref{int3}, describes the coupling to purely spin-3/2 sector of the RS field.
This conclusion is derived both non-perturbatively from the resulting path integral \eref{gipi}, and perturbatively from the calculation of the tree-level amplitude, \Eqref{tree2}.  The gauge-invariant coupling \Eqref{int3}
is thus a good candidate for a consistent cubic interaction of
a scalar, spinor and vector-spinor fields in flat Minkowski space-time.

Some other consistent interactions of the spin-3/2 field can be
immediately written down knowing that they should be restricted
by gauge invariance. For instance,
\bea
\lag_{\pi\De\De}&=& 
g_{\pi\De\De}\, \psib_\mu \ga_5 \tilde{G}^{\mu\nu} \pa_\nu\phi, \\
\lag_{\ga N\De} &=& g_{\ga N\De}\, \Psib \Th_{\al\be,\mu\nu}
G^{\al\be} F^{\mu\nu} + {\rm H.c.},
\eea
where $F^{\mu\nu}$ is the electromagnetic field strength, $\Th$ is a constant
tensor, \eg
\beq
\Th_{\al\be,\mu\nu} = \rg_{\al\mu}\rg_{\be\nu}+ a_1 \rg_{\al\mu}\ga_\be\ga_\nu
+ a_2 \veps_{\mu\nu\al\be} + \mbox{derivative terms},
\eeq
and, finally, $G_{\mu\nu}=\pa_\mu\psi_\nu-\pa_\nu\psi_\mu$,
$\tilde{G}^{\mu\nu}=\veps^{\mu\nu\al\be}\pa_\al\psi_\be.$

An acceptable $\ga\De\De$ interaction  can also be easily found as long
as the coupling to the photon is ``anomalous'', \ie occurs only through
$F^{\mu\nu}$. On the other hand, 
to write down a consistent minimal coupling is not a trivial
task since it is then difficult to satisfy both photon and spin-3/2 gauge
symmetries at the same time. In this case, as well as in other
 cases when one needs to set up lower-derivative interactions,
 supersymmetry might be the only option.

\acknowledgments
I am greatly indebted to M.\ Vasiliev for his advice to consider
the problem using the Hamiltonian path-integral
methods, several subsequent explanations, and comments on this manuscript.
I also would like to thank
O.\ Scholten, J.A.\ Tjon and B.\ de Wit for helpful remarks; 
M.\ Henneaux for a few useful references; and D.V.\ Alhuwalia, 
T.\ Goldman and M.\ Kirchbach for an instructive discussion on
non-local formulations.

The work is supported by {\it de Nederlandse Organisatie voor 
Wetenschappelijk Onderzoek (NWO)} through 
{\it de Stichting voor Fundamenteel Onderzoek der Materie (FOM)}.

\appendix
\section*{St\"uckelberg mechanism for the spin-3/2 field}

Our procedure goes in full analogy to the massive spin-1 case (the Proca model).
We introduce a `St\"uckelberg spinor' $\xi(x)$
replacing $\psi_\mu$ by $\psi_\mu + \pa_\mu\xi$ in the
free Lagrangian \eref{l32}. The Lagrangian reads then as follows,
\beq
\eqlab{stuckl32}
\lag_{\small 3/2}= \half \bar{\psi}_\mu\, \{ \si^{\mu\nu},\,(i\sla{\pa}-m)\}\,\psi_\nu
- m (\pa_\mu\bar{\xi})\si^{\mu\nu}\psi_\nu -m \psib_\mu \si^{\mu\nu}
\pa_\mu\xi,
\eeq
and it is manifestly invariant under the gauge transformation,
\bea
\eqlab{gauget2}
\psi_\mu &\rightarrow& \psi_\mu + \pa_\mu \eps,\no\\
\xi &\rightarrow& \xi - \eps.
\eea
We define the conjugate momenta\footnote{We shall omit
similar formulas for the hermitian-conjugate fields where possible.},
\beq
\pi^{\mu\dagger}(x)= {\pa\lag / \pa \dot{\psi}_{\mu}(x)},\,\,\,\,
\eta^\dagger (x) = {\pa\lag / \pa\dot{\xi}(x)}, 
\eeq
and fundamental Poisson brackets,
\beq
\{\psi_{\mu\tau} (x), \,\pi_\si^{\nu \dagger}(y)\}_P
  =  \de^\nu_\mu\,\de_{\si\tau}\,\de^3(x-y),\,\,\,\,
 \{\xi_\tau (x), \,\eta_\si^{\dagger}(y)\}_P
  =  \de_{\si\tau}\,\de^3(x-y),
\eeq
where $\tau, \si = 0,\ldots,3$ are the spinor indices. We
obtain then the following primary constraints,
\bea
\eqlab{sconstr}
\th_0(x)&=& \pi_0(x),\no\\
\th_i(x)&=&\pi_i(x) - \boxfrac{i}{2} \si_{ij}\psi_j(x)\\
\th_S(x)&=&\eta(x) - m\ga_i\psi_i(x) \no
\eea
and the Hamiltonian,
\beq
H_{\small 3/2}=\int\!\dd^3x\,\ham_{\small 3/2},\,\,\, 
\ham_{\small 3/2} =   \psi_0^\dagger \th_4  
+ \half \psib_i (\veps_{ijk}\, \ga_0\ga_5\,\pa_k + m \si_{ij}) \psi_j
+ m\psib_i \si_{ij}\pa_j\xi + {\rm H.c.},
\eeq
where $\th_4$ is the only secondary constraint, given by
\beq
\th_4(x) = -i \si_{ij}\pa_i\psi_j(x) + m \ga_i
\left(\psi_i(x) + \pa_i\xi(x)\right).
\eeq

We introduce the Dirac bracket with respect to the second-class
constraints $\th_i$ and $\th_S$. Using this bracket the field commutators
take the following form
\bea
\eqlab{dbra}
\{\psi_i(x),\psi^\dagger_j(y)\}_D &=&
 -i(\de_{ij}+ \third\ga_i\ga_j)\,\de^3(x-y),\no\\
\{\xi(x),\xi^\dagger(y)\}_D &=&
 -i\frac{2}{3m^2}\,\de^3(x-y), \\
\{\xi(x),\psi_i^\dagger(y)\}_D &=& \{\psi_i(x),\xi^\dagger(y)\}_D
 = \frac{1}{3m}\ga_i\,\de^3(x-y). \no
\eea
We find then that the secondary constraint commutes with the
Hamiltonian, \ie
\beq
\{\th_4, H_{\small 3/2}\}_D = 0,
\eeq
thus no further constraints arise. We also conclude that
$\th_0$ and $\th_4$ are 
the first-class constraints.

Let us denote the first-class constraints as
$\th^{(1)}=(\th_0,\th_4),$ the corresponding gauge-fixing
conditions as $\vphi=(\vphi_1,\vphi_2),$ and the second-class 
constraints as $\th^{(2)}=(\th_i,\th_S).$ Then, assuming $\vphi$'s
commute among themselves, the path integral
can be put in the following form (see \eg \cite{GiT86,HeT92}),
\bea
\eqlab{genpath}
Z&=&\int\!\DD\psi_\mu\,\DD\psi_\mu^\dagger\,
\DD\xi\,\DD\xi^\dagger\,\DD\pi^\mu
\,\DD\pi^{\mu\dagger}\DD\eta\,\DD\eta^{\dagger}\,
\,\det{\|\{\th^{(1)},\vphi\}_D\|}\,
\left(\det{\|\{\th^{(2)},\th^{(2)}\}_P\| }\right)^{1/2}
\no \\
&&\times \prod \de(\vphi)\, \de(\vphi^\dagger)\,\de(\th)\, \de(\th^\dagger)\,
\exp{\left\{i\int\!\dd^4 x\left[ \pi^{\mu\dagger}\dot{\psi}_\mu
+ \dot{\psi}_\mu^\dagger\pi^\mu + \eta^{\dagger}\dot{\xi}
+ \dot{\xi}^\dagger\eta - \ham_{3/2} \right]\right\} }.
\eea
In our case $\det{\|\{\th^{(2)},\th^{(2)}\}_P\|}$ is just a constant
and can be dropped, since the path integral is defined upto a normalization
factor. 

Clearly one of the gauge-fixing conditions must be proportional
to $\psi_0$ in order to match the $\th_0$ constraint. We take 
$\vphi_1=\psi_0$, then for $\vphi_2$ there is a number of possibilities,
\eg,
\bea
\vphi_2&=&\ga_i\psi_i\,\,\,\, \mbox{(Coulomb gauge)} \no\\
\vphi_2&=&\pa_i\psi_i\,\,\,\,  \mbox{(transverse gauge)} \no\\
\vphi_2&=&\psi_3\,\,\,\,  \mbox{(axial gauge)} \no\\
\vphi_2&=&\xi\,\,\,\,  \mbox{(unitary gauge)} \no
\eea
Let us choose the Coulomb gauge.  
Integrating over the conjugate momenta we then arrive at,
\beq
\eqlab{path3}
Z=\int\!\DD\psi_\mu\,\DD\psi_\mu^\dagger\,
\DD\xi\,\DD\xi^\dagger\,
(\det{[\ga_i\pa_i]})^2
\,\de(\ga_i\psi_i)\,\de(\psi_i^\dagger\ga_i)\,
e^{i\!\int\!\lag_{\small 3/2}}.
\eeq

Now, having the gauge symmetry at our disposal,
we may use the Faddeev-Popov trick \cite{Fad70,FaS90,FaP67} 
to covarianize the measure. We thus obtain
the path integral in a covariant gauge,
\beq
\eqlab{path4}
Z=\int\!\DD\psi_\mu\,\DD\psi_\mu^\dagger\,
\DD\xi\,\DD\xi^\dagger\,
(\det{[\sla\pa]})^2
\,\de(\ga\cdot\psi)\,\de(\psi^\dagger\cdot\ga)\,
e^{i\!\int\!\lag_{\small 3/2}}.
\eeq
Starting from the transverse gauge we would arrive at
\beq
\eqlab{path5}
Z=\int\!\DD\psi_\mu\,\DD\psi_\mu^\dagger\,
\DD\xi\,\DD\xi^\dagger\,
(\det{[\pa_\mu\pa^\mu]})^2
\,\de(\pa\cdot\psi)\,\de(\pa\cdot\psi^\dagger)\,
e^{i\!\int\!\lag_{\small 3/2}}.
\eeq
In these gauges the massless limit can be obtained directly.
On the other hand, taking the unitary gauge and integrating over $\xi$'s
gives us back \Eqref{pi1}.

\end{document}